# Test Sensitivity in the Computer-Aided Detection of Breast Cancer from Clinical Mammographic Screening: a Meta-analysis

Corresponding Author: Jacob Levman[1,2], PhD

1. Institute of Biomedical Engineering

   Department of Engineering Science

   University of Oxford

   Parks Road, Oxford, OX1 3PJ

   United Kingdom

   jacob.levman@eng.ox.ac.uk

   Telephone: 44 01865 617696

2. Sunnybrook Research Institute

   Imaging Research

   University of Toronto

   2075 Bayview Ave.

   Toronto, ON

   M5N 3M5

   Canada

**Abstract**

**Objectives:** To assess evaluative methodologies for comparative measurements of test sensitivity in clinical mammographic screening trials of computer-aided detection (CAD) technologies.

**Materials and Methods:** This meta-analysis was performed by analytically reviewing the relevant literature on the clinical application of computer-aided detection (CAD) technologies as part of a breast cancer screening program based on x-ray mammography. Each clinical study's method for measuring the CAD system's improvement in test sensitivity is examined in this meta-analysis. The impact of the chosen sensitivity measurement on the study's conclusions are analyzed.

**Results:** This meta-analysis demonstrates that some studies have inappropriately compared sensitivity measurements between control groups and CAD enabled groups. The inappropriate comparison of control groups and CAD enabled groups can lead to an underestimation of the benefits of the clinical application of computer-aided detection technologies.

**Conclusions:** The potential for the sensitivity measurement issues raised in this meta-analysis to alter the conclusions of multiple existing large clinical studies is discussed. Two large scale studies are substantially affected by the analysis provided in this study and this meta-analysis demonstrates that computer-aided detection systems are successfully assisting in the breast cancer screening process.

**Keywords:** Computer-aided detection, breast cancer screening, mammography, clinical studies, sensitivity.

**Introduction**

Evidence suggests that the early detection of breast cancer through periodic mammographic screening reduces the mortality associated with the disease [1, 2]. Computer-aided detection (CAD) systems have the potential to improve the breast cancer screening process by marking suspicious tissues as potentially malignant on x-ray mammograms, thus minimizing the likelihood of a cancer being missed by the interpreting radiologist. CAD systems mark suspect cancerous tissues but also incorrectly mark non-malignant tissues, thus although a CAD system may provide improvements in the rate of detection of cancers and may improve the sensitivity of the screening process, it may also cause false positives, leading to higher recall rates and unnecessary biopsies. This article is focused on how clinical CAD studies measure a test's sensitivity and how a clinical study's methodology can affect the study's conclusions.

Research has been ongoing in the design and development of CAD systems to help radiologists with the breast cancer screening process. The research and development of a CAD system typically incorporates rounds of lab based evaluation, comparing the CAD marked results with ground truth data. Once a CAD system has performed successfully in earlier evaluative studies, the technology may reach the stage whereby it is evaluated clinically on ongoing examinations that are being actively relied upon for the detection of breast cancer. Typically, testing a CAD system on an active screening population is performed with commercially available CAD technology. This meta-analysis focuses on those CAD studies that were performed in a clinical setting. The studies included in the detailed analysis herein tested CAD systems in a prospective manner, by measuring the CAD system's performance on a set of mammographic examinations actively being used to screen for breast cancer. Thus this meta-analysis exclusively analyzes those CAD studies that have reached a relatively advanced state of clinical use.

In the course of a typical clinical CAD study, numerous performance metrics are computed to assist with the process of evaluating the performance of the CAD technology being tested (cancer detection rates, test sensitivity, recall rates, biopsy rates, size and stage of detected cancers, test specificity, etc.). Variations exist in the way a clinical study calculates its performance metrics and in particular, variations exist in how those performance metrics are compared. This meta-analysis reviews variations in the methods for measuring a test's sensitivity in clinical CAD studies for breast cancer detection from x-ray mammography and discusses the potential negative effects of relying on a direct comparison of the sensitivity measurements used.

Clinical studies of the effects of CAD technology can be divided into two groups: matched studies and longitudinal studies. In matched studies a radiologist will analyze a mammogram and are then exposed to the results of the CAD technology which may change their diagnosis. Since the imaging examinations are carefully matched and analyzed first without and then with CAD, one can be confident that comparing the measured test sensitivity before and after CAD based screening appropriately reflects the potential performance improvements achieved by the CAD system.

True sensitivities are not computed in these studies as the number of missed cancers is an inherent unknown. Instead, it should be recognized that a sensitivity measurement is relative and as such great care needs to be taken in order to ensure that two measured sensitivity values are indeed appropriate for direct comparison between the control group and the CAD enabled group. Potential problems with a clinical CAD study can occur when the test sensitivities of two groups are compared inappropriately (ie. between CAD enabled screening and screening with no CAD technology).

When a CAD system's relative improvement over non-CAD enabled screening is measured in longitudinal studies it is typically assessed by comparing the test metrics between two large groups (CAD enabled and non-CAD enabled). These groups are not matched, and as such problems can occur when we compare large population groups with a measurement like the test's sensitivity. If a CAD system gets introduced and in its first year of operation increases the yield of cancers detected (ie. a real increase in the cancer detection rate), then it is normal to expect to see an associated increase in the test's measured sensitivity, however this is not necessarily the case. When the breast imaging mammograms are not matched in the study design, then the sensitivity of the control group can be inadvertently inflated relative to the CAD-enabled group. This can occur because missed some cancerous exams in the control group that would have been caught had CAD been used get counted as true negatives when they are in fact false negatives. This makes comparing sensitivity values between the control group and the CAD enabled group potentially misleading. When the control group counts false negatives as true negatives, its sensitivity is artificially inflated relative to the experimental CAD enabled group. This effect is discussed in more detail in the Discussion.

**Materials and Methods**

*Measuring Sensitivity*

A test's sensitivity is typically assessed as the amount of disease detected relative to the total cases of known disease in the population. The typical definition for a test's sensitivity is provided in equation 1.

$$Sensitivity = \frac{TP}{TP + FN} \qquad (1)$$

Where, TP are the true positives: the malignancies caught by the given screening method

FN are the false negatives: cases of known missed cancers

This meta-analysis involved a detailed literature search in order to identify large scale clinical studies looking at the benefits of computer-aided detection enabled breast cancer screening. Each large scale clinical CAD study was analyzed based on the methodology used to compare CAD enabled screening with alternative screening methodologies. In this meta-analysis, each large scale clinical study's test evaluation methodology was analyzed and the potential impact of comparing test sensitivities inappropriately is discussed.

**Results**

The clinical assessment of computer-aided detection technologies for x-ray mammographic breast cancer screening has yielded numerous matched studies [3-14]. Such studies typically incorporate an initial non-CAD enabled reading by a radiologist, followed by reinterpretation with the benefit of the analyzed results produced by the CAD system being tested. These studies are not affected by the arguments made in this meta-analysis as the authors' study design carefully examines individual screening examinations with and without the use of CAD technology. Thus cases where the CAD system detects an otherwise missed tumour are clearly recorded and included in the analysis. Potential benefits of CAD enabled screening are clearly analyzed in these types of studies.

Non-matched population based analyses of CAD technologies account for 7 studies found in the literature [15-21]. Each study is summarized in Table 1, along with the potential extent to which this meta-analysis may contribute to reinterpreting the study's results.

Table 1: CAD studies for x-ray mammography – potential effect of this meta-analysis

| Study | Reported Sensitivity Improvement | Impact of this Meta-analysis |
|---|---|---|
| Gur et al.[15] (JNCI, 2004) | None – other metrics used | none |
| Cupples et al.[16] (AJR, 2005) | None – other metrics used | none |
| Fenton et al.[17] (NEJM, 2007) | 3.6% over pre-CAD | Substantial |
| Gromet[18] (AJR, 2008) | 9% over single reader | Very small |
| Gilbert et al.[19] (NEJM, 2008) | Equal to double reader | Very small |
| James et al.[20] (Radiology, 2010) | Equal to double reader | Very small |
| Fenton et al.[21] (JNCI, 2011) | 1.4% over pre-CAD. CAD had a lower sensitivity compared with non-CAD controls. | Substantial |

**Discussion**

Critical to evaluating a new screening technology is the correct accounting (true positives, false negatives) of those cancers that could be caught by the new CAD-enabled screening process. In a non-matched longitudinal population study, a misleading problem can occur with respect to evaluating the quality of the CAD system. Consider those malignant lesions that potentially can be caught by CAD but are not caught because they are in the control group where CAD wasn't implemented. Cases that can be caught by CAD but are not caught by manual radiological analysis are the most critical cases for accurately assessing the sensitivity improvement of CAD-enabled screening. Non-matched studies that compare test sensitivities between control groups and CAD enabled groups can potentially be misleading due to an assumption of a lack of cancers in the control group's negative screening findings.

In a typical non-matched longitudinal population study, malignant lesions presenting on mammography that are missed by a radiologist but able to be caught by the CAD system and would be caught by future rounds of regular screening are counted as true negatives (ie. they are diagnosed as non-cancerous and then erroneously evaluated as correctly diagnosed). These cases of missed cancers that could have been caught by CAD are critical to assessing the CAD system's performance improvement over standard screening. Instead of being counted as true negatives, these cases should be counted as false negatives (ie. an erroneous non-cancerous diagnosis). An increase in false negatives contributes to lowering the measured test sensitivity (see equation 1). Comparing test sensitivities of a control group with a CAD enabled group in a non-matched study can result in an underestimation of the difference in test sensitivity between the CAD enabled group and the control group. Thus comparing sensitivity measures between two independent groups can imply a smaller sensitivity improvement than was actually accomplished by the introduction of CAD technology because missed cancers in the control group that were CAD detectable are regularly incorrectly counted as correctly diagnosed non-cancers (true negatives) when they are actually incorrectly diagnosed cancers (false negatives).

In preclinical CAD trials, the outlined comparative sensitivity problem is usually not an issue as pre-clinical CAD evaluation does not tend to compare two pools of samples separately. Instead individual exams are typically matched such that multiple screening methods are tested on the exact same mammograms and directly compared. Thus a typical lab-based evaluation of a CAD system does not suffer from the comparative sensitivity problems discussed in this paper. The matched clinical studies investigating the use of CAD [3-14] also do not appear to suffer from the problems with analyzing results by comparing the sensitivity of two different screening methods. In these situations we can be confident that any sensitivity measures produced are a reasonable method by which to compare two screening methodologies.

The longitudinal studies which compare separate populations, one with CAD and one without CAD can lead to erroneous conclusions when analysis of screening technologies is based on comparing sensitivities measured for each group. Two of the six longitudinal CAD studies included in Table 1 in this paper's results avoided the test sensitivity as an evaluative metric and so are not affected by the comparative sensitivity issues presented in this meta-analysis [15, 16]. Three additional longitudinal studies from Table 1 are affected by this comparative sensitivity issue, however, those study also included dual reading in the control group [18-20] which should help minimize the potential negative effects described in this analysis. This is because in the control group, the second reader often identifies extra malignancies that would have been caught by the CAD system, which in turn prevented those cases from being miscounted as true negatives. Thus the expected impact of this meta-analysis on Gromet's study [18], Gilbert's study [19] and James' study [20] is expected to be very small as indicated in Table 1.

Two of the six longitudinal CAD studies are potentially substantially affected by the arguments raised in this analysis [17, 21]. Those two studies' conclusions only emphasized the existence of extremely minor benefits of CAD enabled screening. The effect described in this meta-analysis is liable to have reduced the difference between the measured sensitivities of the control groups relative to the CAD enabled groups because the control groups' sensitivities are not degraded for the missed cancers that would have been caught by CAD had it been deployed in the control population. This may help explain why the more recent study in the *Journal of the National Cancer Institute* [21] reported higher sensitivities in the non CAD-enabled control group relative to the CAD enabled group – an initially surprising finding. This may also explain why only a very small sensitivity improvement was reported when comparing the CAD enabled group with the pre-CAD control group (1.4% improvement). It is interesting to note that although both studies emphasized meager benefits from CAD technology [17, 21], the earlier study in the *New England Journal of Medicine* [17] indicates that CAD screening actually resulted in an increase in the rate of detection of ductal carcinoma in situ (DCIS) by 34% and a decrease in the rate of detection of invasive cancers by 12%, indicating that CAD use may have contributed to shifting the tumour yield towards earlier stage pre-invasive cancers (which are known to have more favourable prognostic characteristics). The later study of the two from the *Journal of the National Cancer Institute* [21] also showed a statistically significant shift towards more DCIS yield in CAD-enabled screening [22], indicating that contrary to the author's conclusions, CAD has exhibited a constructive role in the breast cancer screening process.

In order to help illustrate the problems with comparing measured sensitivities in unmatched longitudinal trials, consider the simple example of a screening center which annually catches 800 cancers with mammography. Furthermore, 200 patients screened at that center are diagnosed with breast cancer annually, even though they had a negative mammogram for a total of 1000 patients diagnosed with breast cancer annually. By the methods used in the literature [17, 21], the center's sensitivity will be 800/(800+200) = 80%. If we then add a CAD system that catches 100 extra tumours in its first round of screening and those tumours would have otherwise been caught in a subsequent round of screening (CAD contributes to catching those tumours earlier), then the sensitivity (as measured [17, 21]) would be 900/(900+200) = 81.8%. Direct comparison reveals just a 1.8% absolute increase in the test sensitivity even though the introduction of the CAD system resulted in an increased yield of malignant tumours of 12.5% (100/800) in the first year of operation.

If this hypothetical screening center had implemented CAD technology one year earlier, then we would expect the CAD system to have yielded an additional 100 lesions in the year it was introduced. The computed sensitivity for the control group prior to implementation of CAD does not account for these missed malignancies that would have been caught had CAD been implemented earlier. Instead those malignancies are missed and erroneously counted as correct diagnoses of exams without malignancies. Consider the control group in a situation where there are 100 missed cancers that could have been caught by CAD and would eventually be caught by a future round of traditional non-CAD based screening. The aforementioned testing methodologies [17, 21] would only account for the normal 200 cases of cancer caught in spite of the implementation of mammographic screening (screened women who present with cancer after a negative mammogram). The control group's sensitivity would be computed as 800/(800+200)=80%, when there are in fact 100 extra missed cancers that were not accounted for and so a more accurate sensitivity for the control population would be 800/(800+300)=72.7%. Comparing this hypothetical control with a CAD enabled group yielding an 81.8% sensitivity demonstrates a solid improvement in the test's sensitivity. Such a comparison is not made however, as these cancers are not accounted for in the computation of the test's sensitivity [17, 21]. It should be noted that the above is just a simple example, however, it clearly illustrates that it is possible for a real improvement in cancer yield of 12.5% (100/800) to result in as little as a 1.8% improvement in sensitivity when relying on problematic methods for comparing sensitivity measurements [17, 21].

The earlier study in the *New England Journal of Medicine* [17] looked exclusively at the performance of a particular CAD technology (ImageChecker, R2 Technology) and averaged its performance across a variety of centers. The more recent study from the *Journal of the National Cancer Institute* [21] averages together 25 CAD screening centers. It was not reported whether all of these centers employed the same commercial CAD technologies. It is expected that different CAD technologies from different vendors will produce different benefits. Furthermore, Dr. Nishikawa and Lorenzo Pesce's analysis demonstrates that different CAD systems for mammographic breast cancer detection can range in disease detection rate improvements from 5 to 20% in cross-sectional trials [23]. Considerable variation also exists between individual radiologists and between different screening centers.

A safer method for evaluating a disease screening technology in a longitudinal clinical context would be to look at the increase in the disease detection rate during the first year/round of screening with a new detection technology. Measuring the disease detection rate beyond this period in a longitudinal trial can lead to misleading conclusions because of the lowered prevalence of disease in the population after introduction of a new more sensitive screening method [23-24]. Additionally, examining the detected tumours' size and stage are appropriate measures for the evaluation of a screening technology.

Sensitivities are relative measurements, when comparing two sensitivities great care must be taken to ensure that the comparison is appropriate so as not to result in misleading conclusions. Of the many large scale clinical studies included in this meta-analysis, the two that are most affected by this paper [17, 21] also demonstrate some of the least benefits from computer-aided detection enabled mammographic screening, indicating that computer-aided detection systems are in fact assisting in mammographic breast cancer screening. The results of this meta-analysis have been published in the journal Radiology [25].

**Acknowledgements**

This work was supported in part by the *Canadian Breast Cancer Foundation*.